\documentclass[seceq,supplement]{ptptex} 
\usepackage{graphicx} 

\title{Overview of $\bar K$-Nuclear Theory and Phenomenology\footnote{ 
Invited talk at the Yukawa International Symposium, 
Kyoto University, December 2006}} 

\subtitle{Search for Narrow Quasibound States} 

\author{Avraham \textsc{Gal}} 

\inst{Racah Institute of Physics, The Hebrew University, 
Jerusalem 91904, Israel}

\abst{Experimental evidence for $\bar K$-nuclear quasibound states is 
briefly reviewed. Theoretical and phenomenological arguments for 
and against deep $\bar K$-nucleus potentials which might allow for 
narrow quasibound states are reviewed, with recent calculations suggesting 
that $\Gamma_{\bar K} \geq 100$ MeV for $B_{\bar K} \leq 100$ MeV. 
Results of RMF calculations that provide a lower limit of 
$\Gamma_{\bar K} \sim 50 \pm 10$ MeV on the width of deeply bound 
states are discussed.} 

\begin{document} 

\maketitle 

\section{Introduction} 

The $\bar K$-nucleus interaction near threshold is strongly attractive and 
absorptive as suggested by fits to the strong-interaction shifts and widths 
of $K^-$-atom levels~\cite{BFG97}. 
Global fits yield `deep' density dependent (DD) optical potentials with 
nuclear-matter depth 
Re~$V_{\bar K}(\rho_0)\sim-(150-200)$~MeV~\cite{FGB93,FGB94,FGM99,MFG06,BFr07}, 
whereas in the other extreme case several studies that fit the low-energy 
$K^-p$ reaction data, including the $I=0$ quasibound state $\Lambda(1405)$ 
as input for constructing self consistently DD optical potentials, obtain 
relatively `shallow' potentials with Re~$V_{\bar K}(\rho_0) \sim -(40-60)$ 
MeV~\cite{SKE00,ROs00,CFG01}. The issue of depth of Re~$V_{\bar K}$ is 
discussed in Section \ref{sec:pot} and the implications of a `deep' potential 
for the existence and properties of $\bar K$-nucleus quasibound states are 
discussed in Section \ref{sec:RMF}. Below we briefly discuss deeply bound 
$K^-$ atomic states and review the debate over deeply bound $\bar K$ nuclear 
states in light nuclei.

\begin{figure}[tbh] 
\centering 
\includegraphics[width=6.5cm]{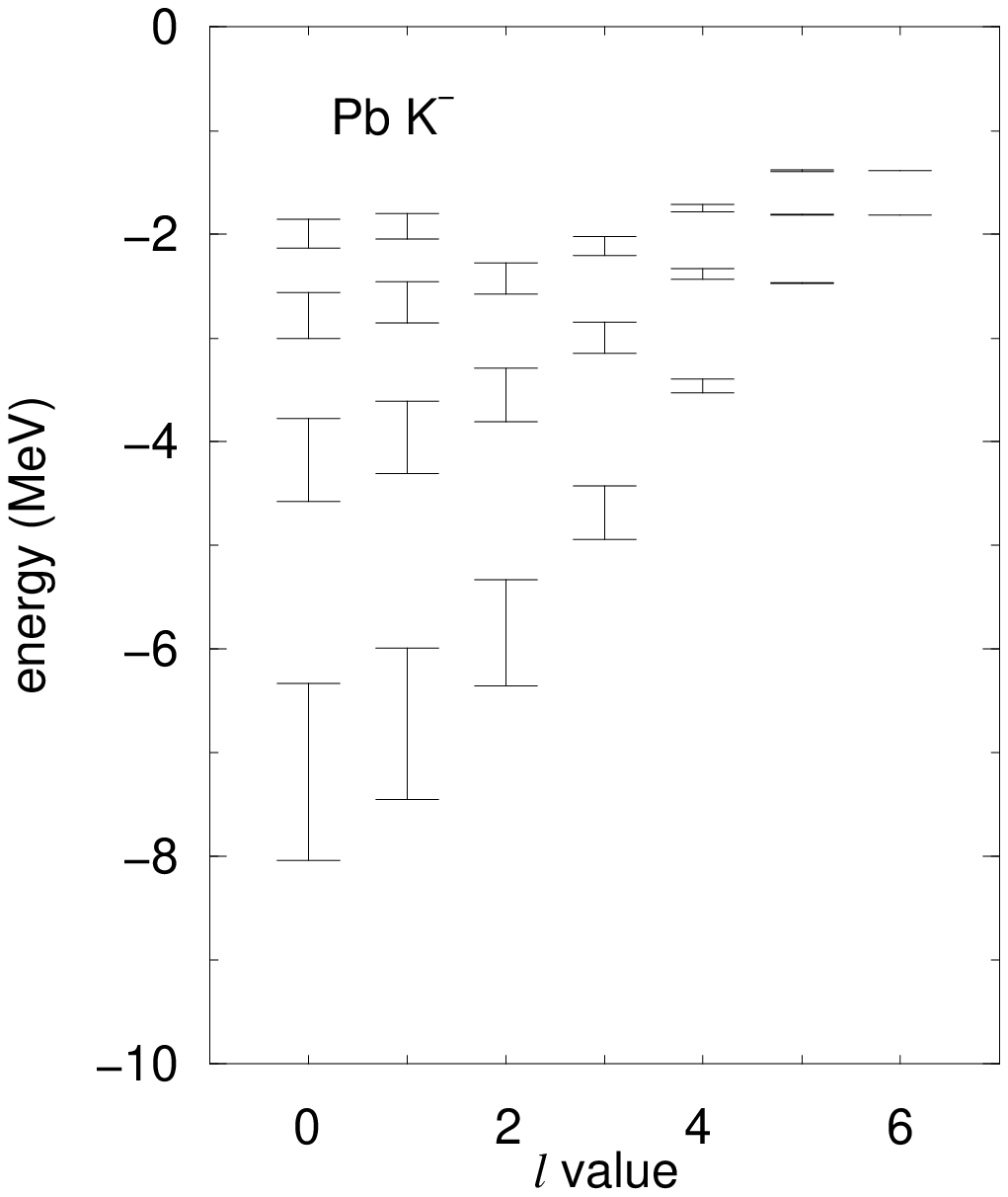} 
\hspace*{3mm} 
\includegraphics[width=6.5cm]{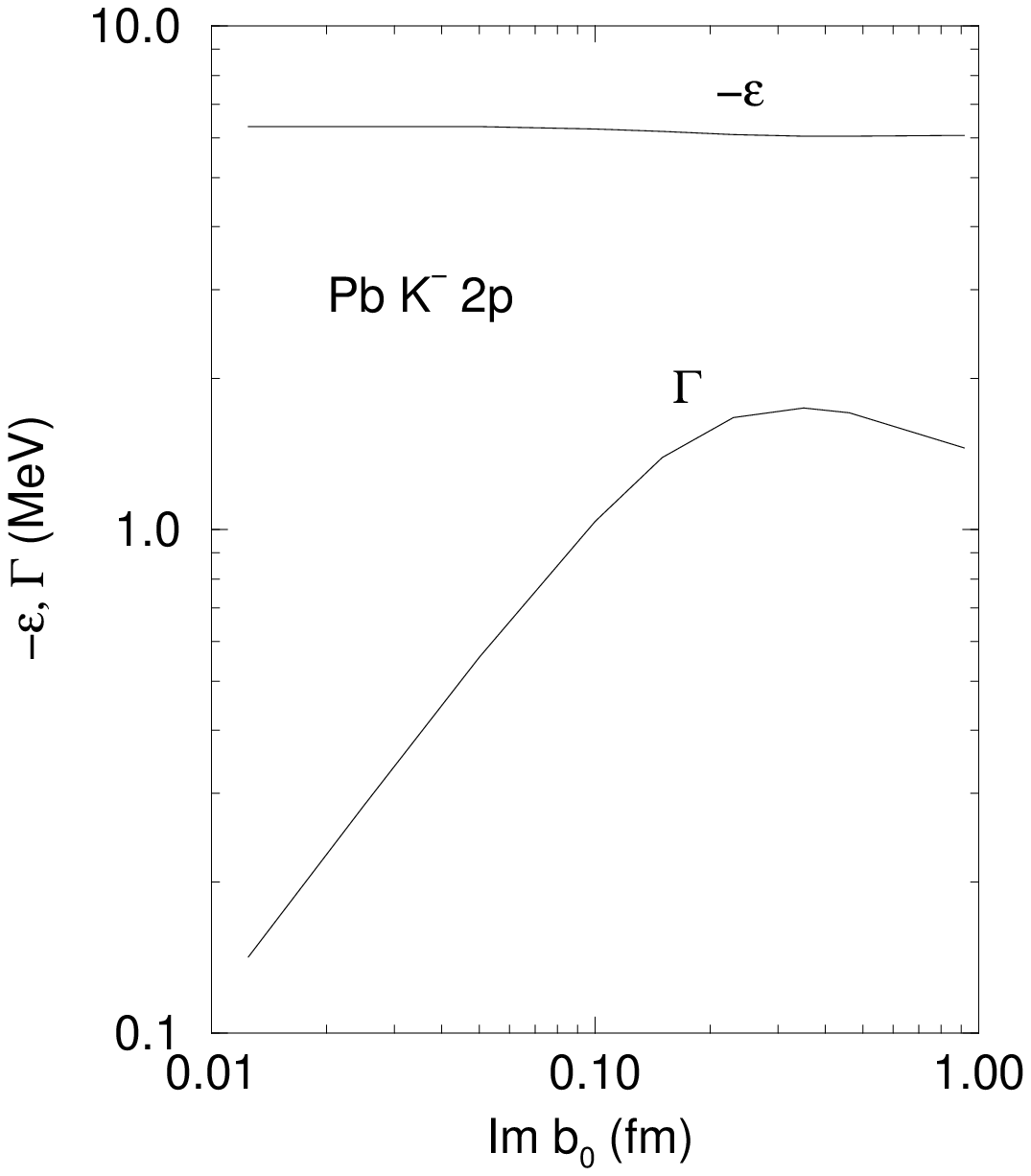} 
\caption{Calculated $K^-$ `deeply bound' atomic states in $^{208}$Pb (left) 
and saturation of width $\Gamma_{\rm atom}$ for the $2p$ `deeply bound' state 
(right) as function of absorptivity, Im~$b_0$, for Re~$b_0 = 0.62$~fm. }
\label{fig:KPb} 
\end{figure} 

\subsection{Deeply bound $K^-$ atomic states} 
\label{sec:atomic db} 
Paradoxically, due to the strong (absorptive) Im~$V_{\bar K}$, 
relatively narrow $K^-$~deeply bound {\it atomic} states are expected  
to exist~\cite{FGa99a}, independently of the size of Re~$V_{\bar K}$.  
Figure~\ref{fig:KPb} from Ref.~\cite{FGa99b} shows on the left-hand side 
a calculated spectrum of $K^-$ atomic states in $^{208}$Pb where, 
in particular, all the circular states below the $7i~(l=6)$ state are 
not populated by X-ray transitions due to the strong $K^-$-nuclear absorption, 
and on the right-hand side it demonstrates saturation of the $2p$ atomic-state 
width for realistically large values of the absorptivity parameter Im~$b_0$ of 
$V_{\bar K}$. The physics behind is that a strong Im~$V_{\bar K}$ 
acts repulsively, suppressing the {\it atomic} wavefunction 
$\Psi_{\rm atom}({\bf r})$ in the region of overlap with Im~$V_{\bar K}$, 
which according to the expression 
\begin{equation} 
\label{eq:width} 
\frac{\Gamma_{\rm atom}}{2} = \frac{\int {\rm Im} (-V_{\bar K}(r))~
|\Psi_{\rm atom}({\bf r})|^2~d{\bf r}}{\int |\Psi_{\rm atom}({\bf r})|^2~
d{\bf r}} 
\end{equation}
implies suppression of the width $\Gamma_{\rm atom}$ as well. The small 
calculated width of `deeply bound' atomic states, 
$\Gamma_{\rm atom} \leq 2$~MeV, also confirmed by the calculation of 
Ref.~\cite{BGN00}, calls for experimental ingenuity to form these levels 
nonradiatively~\cite{FGa99c}. 

\subsection{Deeply bound $K^-$ nuclear states in light nuclei}
\label{sec:nuclear db}
No saturation of $\Gamma_{\bar K}$ holds for $\bar K$-nuclear states which 
retain substantial overlap with the potential. In addition to asking 
(i) whether it is possible at all to bind {\it strongly} $\bar K$ mesons 
in nuclei, one should ask (ii) are such quasibound states sufficiently 
narrow to allow observation and identification? 
The first question was answered affirmatively by Nogami~\cite{Nog63} as 
early as 1963, estimating that the lightest system $K^-pp$ is bound by 
about 10 MeV in its $I=1/2$ state. The first calculation, 
by Yamazaki and Akaishi~\cite{YAk02} 
gave binding energy $B \sim 50$~MeV and width $\Gamma \sim 60$~MeV. 
Preliminary AMD calculations by Dot\'e and Weise~\cite{Wei07,DWe06} 
which implicitly account for $\bar K N - \pi Y$ coupling obtain somewhat 
lower values of $B$ with an estimated width $\Gamma \sim 100$~MeV. 
Coupled-channel $\bar K NN - \pi \Sigma N$ Faddeev 
calculations~\cite{SGM06,ISa07} of $K^-pp$ have confirmed this order of 
magnitude of binding, $B \sim 55 -75$~MeV, differing on the width; 
the calculations of Ref.~\cite{SGM06} give large values 
$\Gamma \sim 100$~MeV for the mesonic width. These Faddeev calculations 
overlook the $\bar K NN \to YN$ coupling to nonmesonic channels which are 
estimated to add, conservatively, 20 MeV to the overall width. Altogether, 
the widths calculated for the $K^-pp$ quasibound state are likely to be 
as large as to make it difficult to identify it experimentally. 

The current experimental and 
theoretical interest in $\bar K$-nuclear bound states was triggered back in 
1999 by the suggestion of Kishimoto~\cite{Kis99} to look for such states 
in the nuclear reaction $(K^{-},p)$ in flight, and by Akaishi and 
Yamazaki~\cite{AYa99,AYa02} who suggested to look for a $\bar K NNN$ $I=0$ 
state bound by over 100 MeV for which the main $\bar K N \to \pi \Sigma$ 
decay channel would be kinematically closed.{\footnote{Wycech had conjectured 
that the width of such states could be as small as 20 MeV~\cite{Wyc86}.}} 
Some controversial evidence for relatively narrow states was presented 
initially in $(K^{-}_{\rm stop},n)$ and $(K^{-}_{\rm stop},p)$ reactions on 
$^4$He (KEK-PS E471)~\cite{ISB03,SBF04} but has just been 
withdrawn (KEK-PS E549/570)~\cite{Iwa06}. $\bar K$-nuclear states were also 
invoked to explain few statistically-weak irregularities in the neutron 
spectrum of the $(K^{-},n)$ in-flight reaction on $^{16}$O (BNL-AGS, parasite 
E930~\cite{KHA03}), but subsequent $(K^{-},n)$ and $(K^{-},p)$ reactions on 
$^{12}$C at $p_{\rm lab}=1$~GeV/c (KEK-PS E548~\cite{Kis06}) have not 
disclosed any peaks beyond the appreciable strength observed below the 
$\bar K$-nucleus threshold. Ongoing experiments by the FINUDA spectrometer 
collaboration at DA$\Phi$NE, Frascati, already claimed evidence for 
a relatively broad $K^- pp$ deeply bound state ($B \sim 115$~MeV) in 
$K^{-}_{\rm stop}$ reactions on Li and $^{12}$C, by observing back-to-back 
$\Lambda p$ pairs from the decay $K^-pp\to\Lambda p$~\cite{ABB05}, but these 
pairs could more naturally arise from conventional absorption processes at 
rest when final-state interaction is taken into account~\cite{MOR06}. 
Indeed, the $K^-_{\rm stop}pn\to \Sigma^- p$ reaction on $^6$Li has also 
been recently observed~\cite{ABB06}. Another recent 
claim of a very deep and narrow $K^- pp$ state ($B \sim 160$~MeV, 
$\Gamma \sim 30$~MeV) is also based on observing decay $\Lambda p$ pairs, 
using $\bar p$ annihilation data on $^4$He from the OBELIX spectrometer 
at LEAR, CERN~\cite{Bre06}. The large value of $B_{K^- pp}$ over 100 MeV 
conjectured by these experiments is at odds with {\it all} the few-body 
calculations of the $K^- pp$ system listed above. Could $\Lambda p$ pairs 
assigned in the above analyses to $K^-pp$ decay in fact result from 
nonmesonic decays of different clusters, say the $\bar K NNN$ 
$I=0$ quasibound state? A definitive identification of the 
$\{\bar K[NN]_{I=1}\}_{I=1/2}$ quasibound state may optimally be reached 
through fully exclusive formation reactions, such as: 
\begin{equation} 
\label{eq:nag}
K^-+^3{\rm He}~ \to ~ n + \{\bar K[NN]_{I=1}\}_{I=1/2,I_z=+1/2},~~~
p + \{\bar K[NN]_{I=1}\}_{I=1/2,I_z=-1/2} \, , 
\end{equation} 
the first of which is scheduled for day-one experiment in J-PARC~\cite{Nag06}. 
Finally, preliminary evidence for a $\bar K NNN$ $I=0$ state with 
$B = 67 \pm 5$~MeV, $\Gamma = 37 \pm 10$~MeV has been recently presented 
by the FINUDA collaboration on $^6$Li by observing back-to-back $\Lambda d$ 
pairs~\cite{ABB07}. It is clear that the issue of $\bar K$ nuclear 
states is far yet from being experimentally resolved 
and more dedicated, systematic searches are necessary.

\section{$\bar K$-nucleus potentials} 
\label{sec:pot} 
\subsection{Chirally motivated models} 
\label{sec:shallow} 

The Born approximation for $V_{\bar K}$ due to the 
leading-order Tomozawa-Weinberg (TW) vector term of the chiral effective 
Lagrangian \cite{WRW97} yields a sizable attraction: 
\begin{equation} 
\label{eq:chiral} 
V_{\bar K}=-\frac{3}{8f_{\pi}^2}~\rho\sim -55~\frac{\rho}{\rho_0}~~{\rm MeV} 
\end{equation} 
for $\rho _0 = 0.16$ fm$^{-3}$, where $f_{\pi} \sim 93$ MeV is the 
pseudoscalar meson decay constant. Iterating the TW term plus 
next-to-leading-order terms, 
within an {\it in-medium} coupled-channel approach constrained 
by the $\bar K N - \pi \Sigma - \pi \Lambda$ data near the 
$\bar K N$ threshold, roughly doubles this $\bar K$-nucleus attraction. 
It is found (e.g. Ref. \cite{WKW96}) that the $\Lambda(1405)$ quickly 
dissolves in the nuclear medium at low density, so that 
the repulsive free-space scattering length $a_{K^-p}$, as function of 
$\rho$, becomes {\it attractive} well below $\rho _0$. 
Since the attractive $I=1$ $a_{K^-n}$ is only weakly density 
dependent, the resulting in-medium $\bar K N$ isoscalar scattering length 
$b_0(\rho)={\frac{1}{2}}(a_{K^-p}(\rho)+a_{K^-n}(\rho)$) translates into 
a strongly attractive $V_{\bar K}$: 
\begin{equation} 
\label{eq:trho} 
V_{\bar K}(r) = -{\frac{2\pi}{\mu_{KN}}}~b_0(\rho)~\rho(r)~, 
~~~~{\rm Re}~V_{\bar K}(\rho_0) \sim -110~{\rm MeV}\,. 
\end{equation} 
However, when $V_{\bar K}$ is calculated {\it self consistently}, 
namely by including $V_{\bar K}$ in the propagator $G_0$ used in the 
Lippmann-Schwinger equation determining $b_0(\rho)$, one obtains 
Re~$V_{\bar K}(\rho_0)\sim -(40 - 60)$~MeV~\cite{SKE00,ROs00,CFG01}. 
The main reason for this weakening of $V_{\bar K}$, 
approximately back to that of Eq.~(\ref{eq:chiral}), 
is the strong absorptive effect which $V_{\bar K}$ exerts within $G_0$ to 
suppress the higher Born terms of the $\bar K N$ TW potential. 

\begin{figure}[tbh] 
\centering 
\includegraphics[width=6.8cm]{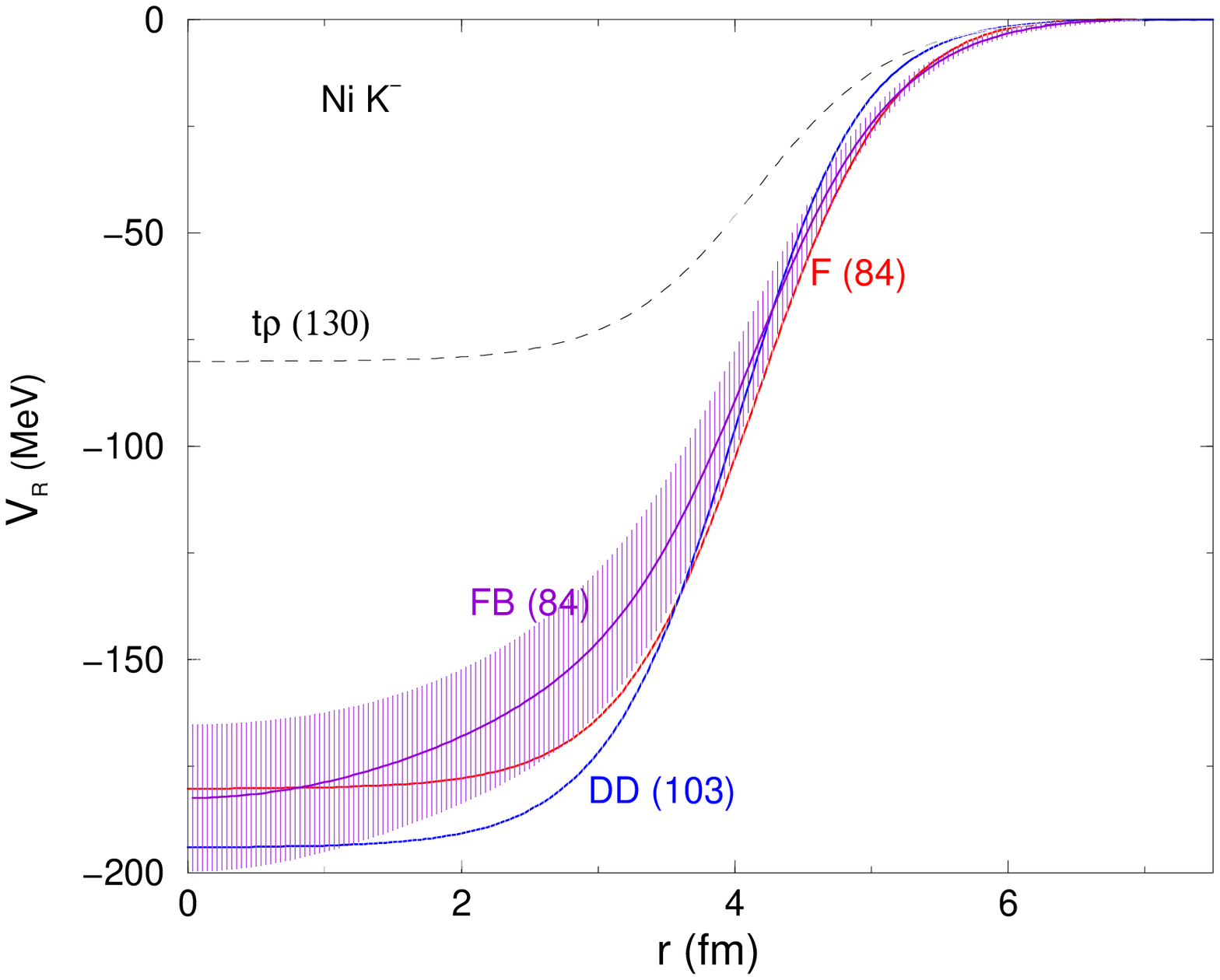} 
\hspace*{3mm} 
\includegraphics[width=6.5cm]{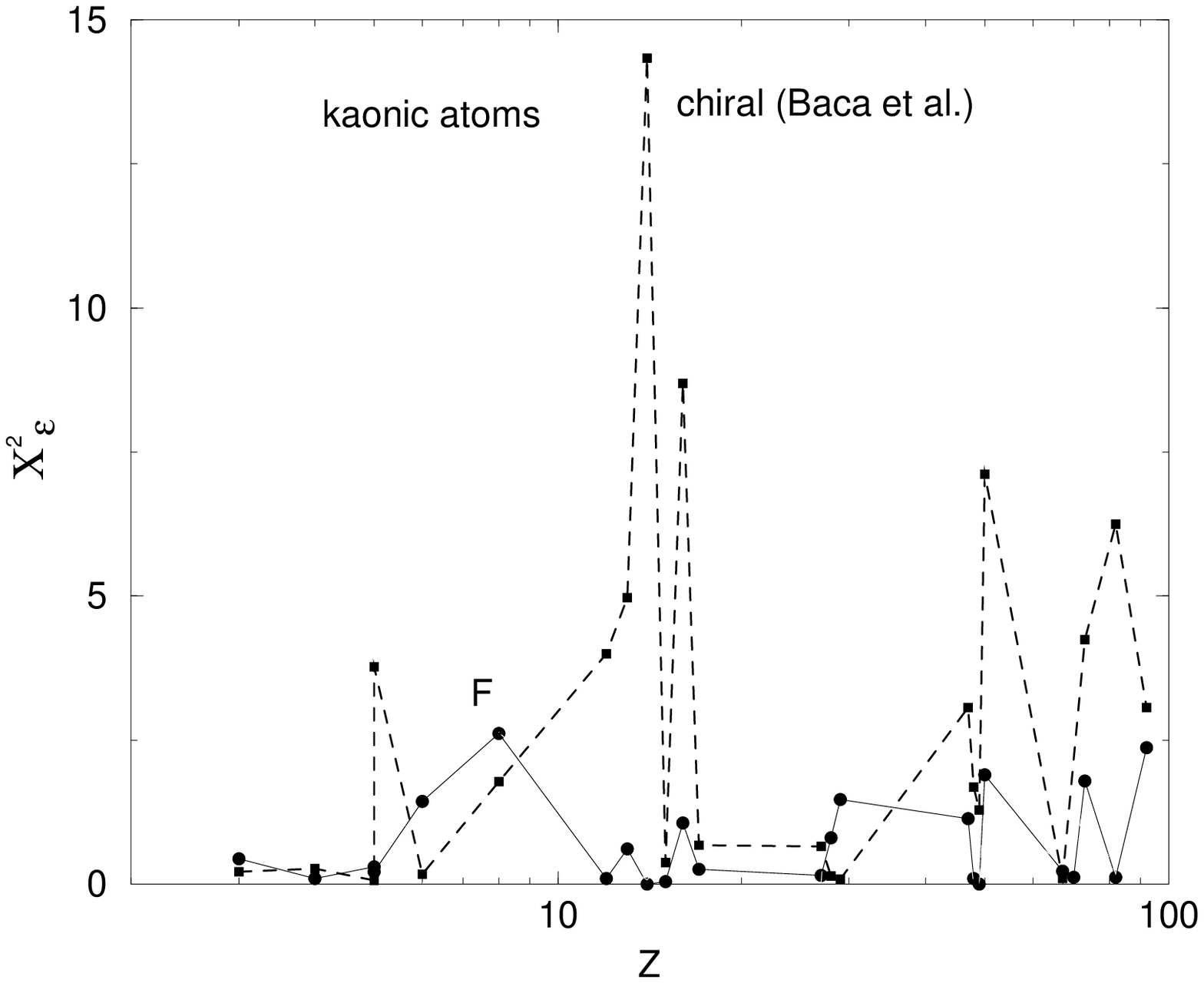}
\caption{Real part of the $\bar K-^{58}{\rm Ni}$ potential obtained 
in a global fit to $K^-$-atom data using the model-independent FB 
technique~{\protect \cite{BFr07}}, in comparison with other best-fit 
potentials and $\chi^2$ values in parentheses (left), and contributions 
to the $\chi^2$ of $K^-$ atomic shifts for the {\it deep} potential F 
from Ref.~{\protect \cite{MFG06}} and for the {\it shallow} chirally-based 
potential from Ref.~{\protect \cite{BGN00}} 
(right). } 
\label{fig:atom} 
\end{figure}

\subsection{Fits to $K^-$-atom data} 
\label{sec:deep} 

The $K^-$-atom data used in global fits~\cite{BFG97} span a range of nuclei 
from $^7$Li to $^{238}$U, with 65 level-shift, -width and -yield data points. 
Figure~\ref{fig:atom} shows on the left-hand side fitted ${\rm Re}~V_{\bar K}$ 
for $^{58}$Ni, for a $t\rho$ potential with a complex strength $t$, for two 
density-dependent potentials~\cite{MFG06} marked by DD and F, and for 
a {\it model independent} potential obtained by adding a Fourier-Bessel (FB) 
series to a $t\rho$ potential.~\cite{BFr07} The depth of the $t\rho$ potential 
is about $70-80$~MeV, whereas the density-dependent potentials (including FB) 
are considerably deeper, $150-200$~MeV. Although DD and F have very different 
parameterizations, the resulting potentials are quite similar to each other. 
These latter potentials yield substantially lower $\chi ^2$ values (shown in 
the figure) than the $\chi ^2$ value for the $t\rho$ potential. 
The superiority of the {\it deep} potential F to a {\it shallow} potential 
based on the self-consistent chiral model of Ramos and Oset~\cite{ROs00} 
in reproducing the $K^-$-atom level shifts is demonstrated on the right-hand 
side of Fig.~\ref{fig:atom}. In particular, the shape of potential F departs
appreciably from  $\rho (r)$ for $\rho (r)/\rho_0 \leq 0.2$, where the
physics of the $\Lambda(1405)$ plays a role. It is a challenge for 
chiral-model practitioners to match the quality of potential F's atomic fit. 

\subsection{Further considerations}
\label{sec:add}


(i) QCD sum-rule estimates~\cite{Dru06} for vector (v) and scalar (s) 
self-energies give: 
\begin{eqnarray} 
\label{eq:QCDv} 
\Sigma_v(\bar K) &\sim & -\frac{1}{2}~\Sigma_v(N)~\sim~
-\frac{1}{2}~(200)~{\rm MeV}~ =~-100~{\rm MeV}\,,\\ 
\Sigma_s(\bar K) &\sim & \frac{m_s}{M_N}~\Sigma_s(N)~\sim~
\frac{1}{10}~(-300)~{\rm MeV}~ =~ -30~{\rm MeV}\, ,
\label{eq:QCDs}
\end{eqnarray}
where $m_s$ is the strange-quark (current) mass. The factor 1/2 in 
Eq.~(\ref{eq:QCDv}) is due to the one nonstrange antiquark $\bar q$ in the 
$\bar K$ out of two possible, and the minus sign is due to $G$-parity going 
from $q$ to $\bar q$. This rough estimate gives then 
$V_{\bar K}(\rho_0) \sim -130$~MeV. The QCD sum-rule approach essentially 
refines the mean-field argument~\cite{SGM94,BRh96} 
\begin{equation} 
\label{eq:MF} 
V_{\bar K}(\rho_0)~\sim~\frac{1}{3}~(\Sigma_s(N)-\Sigma_v(N))~\sim~
-170~{\rm MeV}\,,
\end{equation} 
where the 1/3 factor is again due to the one nonstrange antiquark in the 
$\bar K$, but here with respect to the three nonstrange quarks of the nucleon. 

(ii) The ratio of $K^-/K^+$ production cross sections in nucleus-nucleus and
proton-nucleus collisions near threshold, measured by the Kaon spectrometer 
(KaoS) collaboration at SIS, GSI, gives clear evidence for a strongly 
attractive $V_{\bar K}$, estimated~\cite{SBD06} as 
$V_{\bar K}(\rho_0) \sim -80$~MeV by relying on BUU transport calculations 
normalized to the value $V_K(\rho_0) \sim +25$~MeV. 
Since $\bar K NN \to YN$ absorption apparently was disregarded in these 
calculations, a deeper $V_{\bar K}$ may follow when it is included.

\section{RMF dynamical calculations} 
\label{sec:RMF} 

\subsection{$\bar K$-nucleus RMF methodology} 

\begin{figure}[tbh]
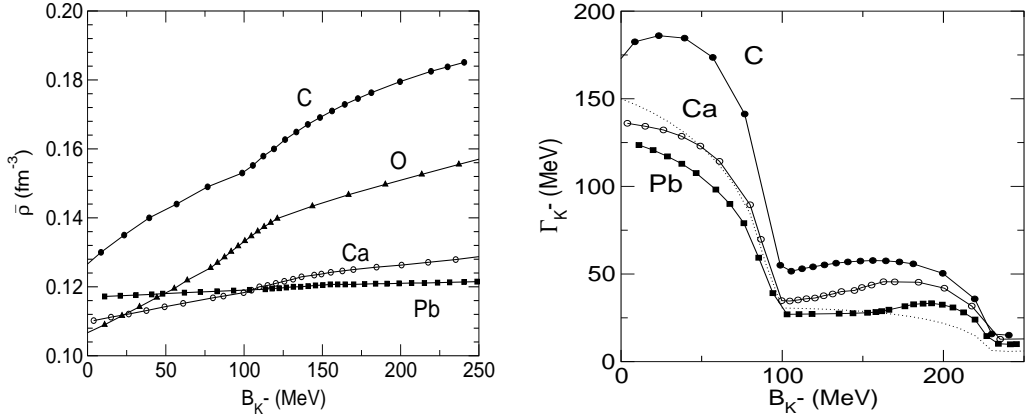
 
\centering 
\includegraphics[height=5.5cm,width=6.5cm]{K05fig6.eps} 
\hspace*{3mm} 
\includegraphics[height=5.5cm,width=6.5cm]{K06fig4.eps} 
\caption{Dynamically calculated average density $\bar \rho$ (left) and 
widths $\Gamma_{K^-}$ (right) of $1s$ $K^-$-nuclear states in the nuclei 
denoted, as function of the $1s$ $K^-$ binding energy.~\cite{MFG06} }
\label{fig:rho} 
\end{figure}

In this model, expanded in Ref. \cite{MFG06}, the (anti)kaon interaction with 
the nuclear medium is incorporated by adding to ${\cal L}_N$ the Lagrangian 
density ${\cal L}_K$: 
\begin{equation}
\label{eq:Lk}
{\cal L}_{K} = {\cal D}_{\mu}^*{\bar K}{\cal D}^{\mu}K -
m^2_K {\bar K}K
- g_{\sigma K}m_K\sigma {\bar K}K\; .
\end{equation} 
The covariant derivative
${\cal D_\mu}=\partial_\mu + ig_{\omega K}{\omega}_{\mu}$ describes
the coupling of the (anti)kaon to the vector meson $\omega$.
The(anti)kaon coupling to the isovector $\rho$ meson was found to have 
negligible effects. 
The $\bar K$ meson induces additional source terms in the equations of motion 
for the meson fields $\sigma$ and $\omega_0$. It thus affects the scalar 
$S = g_{\sigma N}\sigma$ and the vector $V = g_{\omega N}\omega_0$ potentials 
which enter the Dirac equation for nucleons, and this leads to rearrangement 
or polarization of the nuclear core, as shown on the left-hand side of 
Fig.~\ref{fig:rho} for the calculated average nuclear density 
$\bar \rho = \frac{1}{A}\int\rho^2d{\bf r}$ 
as a function of $B_{K^-}$ for $K^-$ nuclear $1s$ states across the periodic 
table. It is seen that in the light $K^-$ nuclei, $\bar \rho$ increases 
substantially with $B_{K^-}$ to values about 50\% higher than without the 
$\bar K$.{\footnote{The increase of the central nuclear densities is bigger, 
up to 100\%, and is nonnegligible even in the heavier $K^-$ nuclei where it 
is confined to a small region of order 1~fm.}} 
Furthermore, in the Klein-Gordon 
equation satisfied by the $\bar K$, the scalar $S = g_{\sigma K}\sigma$ and 
the vector $V = -g_{\omega K}\omega_0$ potentials become 
{\it state dependent} through the {\it dynamical} density dependence of the 
mean-field potentials $S$ and $V$, as expected in a RMF calculation. 
An imaginary ${\rm Im}~V_{\bar K} \sim t\rho$ was added, fitted to the 
$K^-$ atomic data~\cite{FGM99}. It was then suppressed by an energy-dependent 
factor $f(B_{\bar K})$, considering the reduced phase-space for the initial 
decaying state and assuming two-body final-state kinematics for the decay 
products in the $\bar K N \to \pi Y$ mesonic modes ($80\%$) and in the 
$\bar K NN \to Y N$ nonmesonic modes ($20\%$).  

The RMF coupled equations were solved self-consistently. For a rough idea, 
whereas the static calculation gave $B_{K^-}^{1s} = 132$~MeV
for the $K^-$ $1s$ state in $^{12}$C, using the values 
$g^{\rm atom}_{\omega K},~g^{\rm atom}_{\sigma K}$ corresponding to the 
$K^-$-atom fit, the dynamical calculation gave $B_{K^-}^{1s} = 172$~MeV. 
In order to scan a range of values for $B_{K^-}^{1s}$, $g_{\sigma K}$ and 
$g_{\omega K}$ were varied in given intervals of physical interest. 

\subsection{Binding energies and widths}

Beginning approximately with $^{12}$C, the following conclusions may be drawn: 
(i) For given values of $g_{\sigma K},g_{\omega K}$, the $\bar K$ binding 
energy $B_{\bar K}$ saturates, except for a small increase due to the 
Coulomb energy (for $K^-$). 
(ii) The difference between the binding energies calculated dynamically and 
statically, $B_{\bar K}^{\rm dyn} - B_{\bar K}^{\rm stat}$, is substantial 
in light nuclei, increasing with $B_{\bar K}$ for a given value of $A$, and 
decreasing monotonically with $A$ for a given value of $B_{\bar K}$. 
It may be neglected only for very heavy nuclei. The same holds for the 
nuclear rearrangement energy $B_{\bar K}^{\rm s.p.} - B_{\bar K}$ which is 
a fraction of $B_{\bar K}^{\rm dyn} - B_{\bar K}^{\rm stat}$. 
(iii) The width $\Gamma_{\bar K}(B_{\bar K})$ decreases monotonically 
with $A$, as shown in the right-hand side of Fig.~\ref{fig:rho} 
\begin{figure}[tbh] 
\centering 
\includegraphics[width=6.5cm,angle=-90]{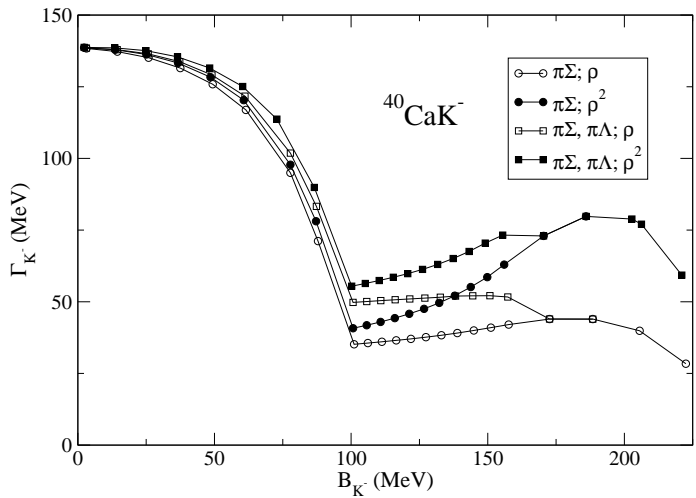} 
\caption{Dynamically calculated widths of the $1s$ $K^-$-nuclear 
state in $^{~~40}_{K^-}$Ca for $\pi \Sigma + \pi \Lambda$ vs. $\pi \Sigma$ 
final mesonic absorption channels, and for $\rho^2$ vs. $\rho$ 
dependence of the final nonmesonic absorption channels, 
as function of the $K^-$ binding energy, from Gazda et al.~\cite{MFG06} }
\label{fig:Gamma}  
\end{figure} 
for $1s$ states. The dotted line shows the static `nuclear-matter' limit 
corresponding to the $K^-$-atom fitted value ${\rm Im}~b_0=0.62$ fm and for 
$\rho(r)=\rho_0=0.16$ fm$^{-3}$, using the same phase-space suppression factor 
as in the `dynamical' calculations. It is clearly seen that the functional 
dependence $\Gamma_{K^-}(B_{K^-})$ follows the shape of the dotted line. 
This dependence is due primarily to the binding-energy
dependence of $f(B_{K^-})$ which falls off rapidly until
$B_{K^-} \sim 100$~MeV, where the dominant
$\bar K N \rightarrow \pi \Sigma$ gets switched off, and then stays
rather flat in the range $B_{K^-} \sim 100 - 200$~MeV where the width is 
dominated by the $\bar K NN \to YN$ absorption modes. The widths
calculated in this range are considerably larger than given by the dotted 
line (except for Pb in the range $B_{K^-} \sim 100 - 150$~MeV) due 
to the dynamical nature of the RMF calculation, whereby the nuclear density 
is increased by the polarization effect of the $K^-$. 
Adding perturbatively the residual width neglected in this calculation, 
partly due to the $\bar K N \to \pi \Lambda$ secondary mesonic decay channel, 
a lower limit $\Gamma_{\bar K} \geq 50 \pm 10$~MeV is obtained for the 
binding energy range $B_{K^-} \sim 100 - 200$~MeV.
Figure~\ref{fig:Gamma} shows the effect of splitting the $80\%$ mesonic decay 
width, previously assigned to $\pi \Sigma$, between $\pi \Sigma$ ($70\%$) and 
$\pi \Lambda$ ($10\%$), and also 
of simulating the $20\%$ nonmesonic absorption channels by a $\rho^2$ 
dependence compared to ${\rm Im}~V_{\bar K} \sim t\rho$ used by Mare\v{s} 
et al.~\cite{MFG06} These added contributions make the above lower limit 
$\Gamma_{\bar K} \geq 50 \pm 10$~MeV a rather conservative one.

\section{Conclusions and acknowledgements} 

I have reviewed the phenomenological and theoretical evidence for 
a substantially attractive $\bar K$-nucleus interaction potential, from 
a `shallow' potential of depth $40-60$ MeV to a `deep' potential of depth 
$150-200$ MeV at nuclear-matter density. In particular, I demonstrated the 
considerable extent to which the {\it deep} potentials are supported by 
$K^-$ atomic data fits. I then reported on recent 
{\it dynamical} calculations~\cite{MFG06} for deeply quasibound 
$\bar K$-nuclear states across the periodic table. Substantial polarization 
of the core nucleus was found in light nuclei, but the `high' densities 
reached are considerably lower than those found in the few-body calculations 
due to Akaishi, Yamazaki and collaborators~\cite{YAk02,AYa02,DHA04}. 
The width $\Gamma_{\bar K}$ of quasibound states, as function of the binding 
energy $B_{\bar K}$, reflects the decay phase-space suppression on top of the 
increase provided by the density of the compressed nuclear core, resulting in 
a lower limit $\Gamma_{\bar K} \geq 50 \pm 10$~MeV, particularly useful for 
$B_{\bar K} \geq 100$ MeV.

I wish to thank Eli Friedman, Ji\v{r}\'{\i} Mare\v{s}, Nina Shevchenko and 
Wolfram Weise for stimulating discussions, and the organizers 
of the 2006 Yukawa International Symposium on ``New Frontiers in QCD" at 
Kyoto University, particularly Teiji Kunihiro (Chair), Atsushi Hosaka, 
Daisuke Jido and Tetsuo Hyodo for their helpful assistance. I am grateful 
to Toshio Motoba and to Makoto Oka for their kind hospitality within the 
JSPS Fellowship program. This work is supported in part by the Israel 
Science Foundation, Jerusalem, grant 757/05.

\end{document}